\documentclass[12pt]{article}
\usepackage{graphicx,epsfig,makeidx}

\author{ S. Galam$^{1,2}$ and A. Mauger$^{1,3}$
\\ $^{1}$Laboratoire des Milieux D\'esordonn\'es et H\'et\'erog\`enes\\
Case 86, 4 place Jussieu, F-75252 Paris Cedex 05\\
   $^{2}$Centre de Recherche en \'Epist\'emologie Appliqu\'ee, \\
CREA - \'Ecole Polytechnique,
CNRS UMR 7656, \\1,  rue Descartes, 75005 Paris \\
 $^{3}$F\'{e}d\'{e}ration de Recherche Mati\`{e}re et
Syst\`{e}mes Complexes, \\CNRS UMR,
140, rue de Lourmel, Paris 75015 }

\title{Possible crossover of a non universal quantity at the upper
critical dimension}\date{(galam@ccr.jussieu.fr; mauger@ccr.jussieu.fr}

\begin{document}

\maketitle

\begin{abstract}

We report on a possible crossover of a non universal quantity
at the upper critical dimensionality in the field of 
percolation.
Plotting recent estimates for site percolation thresholds
of hypercubes in dimension $6\le d\le 13$ against corresponding 
predictions from the GM formula
$p_c=p_0[(d-1)(q-1)]^{-a}d^b$ for percolation thresholds, a significant
departure of $p_c$ is observed for $d\geq 6$. This result is 
reminiscent of the crossover undergone by universal quantities in 
critical phenomena. For bond percolation, the evidence of such a crossover of dimensionality would require an improvement of the GM formula to reach a relative error  of typically $0.2\%$, while it is currently at $0.9\%$ for hypercubes.
\end{abstract}

{\em PA Classification Numbers:\/}  64.60.Ak, 05.70.Jk\\[6ex]
\newpage

\section{Introduction}
The discovery of renormalization group technics by Wilson in the early
seventies has allowed the powerful elucidation of the mystery of
critical phenomena \cite{wilson}. It is based on the existence of
relevant variables,
irrelevant variables and universality classes. Accordingly all
parameters are classified as  universal quantities and non universal
quantities. For continuous  phase transitions the critical exponents are
universal while critical  temperatures are not.

In this framework dimension play a key role
to categorize the effects of fluctuations. At very low dimensions,
fluctuations are too strong and prevent any long range order to
occur. The limit from which  it does not happen is called the lower
critical dimension $d_l$.  Only for  $d>d_l$ can long range order
sustain fluctuations. On the other extreme, there exists some
dimension $d_c$ called the upper critical dimension $d_c$, beyond
which fluctuations are averaged out and do not influence the critical
properties. For $d>d_c$ there exists only one class of universality:
the mean field one. Therefore it is in the range $d_l<d<d_c$ that
fluctuations are instrumental to determine the critical properties.

In
parallel, percolation is a geometric phenomena with no temperature.
However it was shown to be indeed identical to usual critical
phenomena with $d_c=6$. Therefore its  critical exponents are
universal quantities while percolation thresholds $p_c$ are not.
Accordingly the value of $p_c$ must be calculated for each system and
varies from one geometry to another. However, at odd with this
universality principle, a lot of efforts have been devoted to the
finding of  formulas for percolation threshold since about
half a century. Several
formulas have been proposed, which involve only two parameters : the
dimension
$d$ and the coordination number $q$ \cite{vysso,sahi,gal1,
gal2,gm,gal3}. The limits of such a choice has been discussed by J. C.
Wierner et al
\cite{wier1,wier2}, who also considered that the most accurate of
such formulas are the Galam-Mauger (GM) laws \cite{gm}. Indeed the
high degree of accuracy of the GM law predictions hints for the
existence of an underlying universality principle for percolation
thresholds.

In this work, we report on a possible crossover
of a non universal quantity at the upper critical dimensionality in
the field of percolation. Using the GM law and a series of recent numerical
estimates for hypercube percolation thresholds ( $6\le d\le 13$), 
site percolation thresholds are found  to
undergo a drastic change of behavior  at the percolation upper 
critical dimension
$d_c=6$. This result is reminiscent of the crossover undergone by 
universal quantities in critical phenomena. At contrast, nothing 
similar is
evidenced at any dimension for the bond percolation.

Considering only the case
$d>2$ of interest in the present work, the GM law is split in two. One
applies to
$3\le d\le d_c$, and can be written:
\begin{equation}
p_c=p_0[(d-1)(q-1)]^{-a}d^b\qquad 3\le d\le d_c.
\end{equation}
The site percolation
$p_c^S$ is approximated by Eq. (1), with
$b=0,
\, p_0=1.2868,\,a=0.6160$. As for the bond percolation
threshold, $p_0=0.7541,\, b=a=0.9346$. This law corresponds to the
so-called second class (the first one being related to $d=2$
only), and will be called for this reason the GM2 law. Eq. (1) cannot
extend up to
$d\rightarrow
\infty$. Among several reasons outlined in \cite{gm}, one comes from the
fact that, in this limit, the percolation threshold should reduce to that
of the Cayley tree for both sites and bonds. In other words, one must
recover the Bethe asymptotic limit:
\begin{equation}
p_c^S=p_c^B=(q-1)^{-1} \qquad d\rightarrow\infty.
\end{equation}
Eq. (2) is violated by the GM2 law. This drawback of the GM2 law
was the main motivation for introducing another law
associated to a third class, which applies at high
dimension, and has the proper Bethe asymptotic limit. This is the
asymptotic GM3 law:
\begin{equation}
p_c=2^{a-1}[(d-1)(q-1)]^{-a}d^{2a-1},\qquad d\gg d_c
\end{equation}
with $a=0.08800$ for sites, and $a=0.3685$ for bonds.

Despite the fact that the GM2 law is not exact, its accuracy is
sufficient to materialize the dimensional dependence of the percolation
threshold for the bcc, fcc, hypercubic lattices up to $d=6$. Due to the
lack of data for the percolation thresholds in larger dimension,
however, it was not possible so far to explore the existence of a critical
crossover dimension
$d_c$ above which the percolation threshold would follow a
formula different from the GM2 law (approximated by he GM3
law). Recent Monte Carlo estimates for site and bond
percolation thresholds with negligible standard deviations in simple
hypercube lattices from $d=6$ up to $d=13$ \cite{grass} now makes this
investigation possible.

It is the purpose of this work to make the
comparison of the data including these Monte Carlo results with the
predictions of Eq. (1) and (3).
For site percolation
threshold, the crossover between the two laws is clearly evidenced, at
dimension
$d_c= 6$. For bond percolation thresholds, however, no sizeable deviations
from the GM2 law as defined by Eq. (1) is detected, up to
the highest dimension
$d=13$ investigated. Such a crossover for bonds, if it exists, cannot be
detected, since the GM2 and GM3 laws do not depart significantly from
each other in the range $7\le d\le 13$.

\section{Analysis}
The numerical estimates $p_c^S$, $p_c^B$ of the percolation thresholds for 
sites and bonds, respectively, are
reported in Table 1, together with the results of the GM2 and GM3 laws.
For sc lattices at
d=5 and d=6, the data of refs. \cite{stau,gau} have been substituted by
those of ref.\cite{grass}, since they are more accurate; we shall return
to this point later on.

   \begin{table}
\label{tbl}
\begin{center}
   \hspace*{-0.80cm}
\begin{tabular}{|c|c|c|c||c|c|c|}
\hline
\hline
\multicolumn{1}{|c|}{lattice}
&\multicolumn{1}{c}{$p_c^S(GM2)$}
&\multicolumn{1}{c}{$p_c^S(GM3)$}
&\multicolumn{1}{c||}{$p_c^S (nu)$}
&\multicolumn{1}{c}{$p_c^B(GM2)$}
&\multicolumn{1}{c}{$p_c^B(GM3)$}
&\multicolumn{1}{c|}{$p_c^B (nu)$} \\
\hline
Kagom\'e & 0.65400 & 0.59264 & 0.65270 & 0.51620 & 0.38885 & 0.52440 \\
Diamond & 0.42675 & 0.43824 & 0.43000 & 0.39454 & 0.24984 & 0.38800 \\
sc (d=3) & 0.31154 & 0.27957 & 0.31160 & 0.24476 & 0.20697 & 0.24880 \\
bcc (d=3) & 0.25322 & 0.20792 & 0.24600 & 0.17872 & 0.18284 & 0.18030 \\
fcc (d=3) & 0.19168 & 0.13969 & 0.19800 & 0.11714 & 0.15479 & 0.11900 \\
sc (d=4) & 0.19725 & 0.18109 & 0.19700 & 0.16009 & 0.14599 & 0.16010 \\
fcc (d=4) & 0.094794 & 0.063570 & 0.098000 & & & \\
sc (d=5) & 0.14152 & 0.13352 & 0.14100 & 0.11917 & 0.11287 & 0.11820 \\
fcc (d=5) & 0.057352 & 0.036740 & 0.054000 & & & \\
sc (d=6) & 0.10901 & 0.10562 & 0.10902 & 0.095092 & 0.092028 & 0.094202\\
\cline{2-4}
sc (d=7) & 0.087898 & 0.087319 & 0.088951 & 0.079233 & 0.077697 &
0.078675 \\
sc (d=8) & 0.073191 & 0.074401 & 0.075210 & 0.067991 & 0.067232 &
0.067708 \\
sc (d=9) & 0.062410 & 0.064802 & 0.065210 & 0.059602 & 0.059255 &
0.059496 \\
sc (d=10) & 0.054198 & 0.057390 & 0.057593 & 0.053097 & 0.052971 &
0.053093 \\
sc (d=11) & 0.047756 & 0.051497 & 0.051590 & 0.047904 & 0.047893 &
0.047795 \\
sc (d=12) & 0.042578 & 0.046699 & 0.046731 & 0.043659 & 0.043704 &
0.043724 \\
sc (d=13) & 0.038336 & 0.042718 & 0.042715 & 0.040124 & 0.040190 &
0.040188 \\
\hline
\hline
\end{tabular}
\caption{ Numerical estimates for the percolation
thresholds $p_c^S (nu)$, together with the results of the GM2 and GM3
laws. For sc lattices at d=5 and d=6, the data of refs. \cite{stau,gau}
have been substituted by
those of ref.\cite{grass}, since they are more accurate.}
\end{center}
\end{table}

\subsection{site percolation}
   As $b=0$ for
site percolation, the GM2 law is best illustrated in a log-log plot of
$p_c^S$ versus
$(d-1)(q-1)$ in which case it is a straight line. This is illustrated in
Fg. (1). We have reported on the same plot the numerical
results of $p_c^S$ taken from \cite{grass,stau,gau}.  For comparison,
we have also reported (crosses)
the values of
$p_c^S$ predicted by the asymptotic GM3 law. Note according the GM3 law,
$(d-1)(q-1)$ is not the pertinent variable, hence a random-like
distribution of the crosses which cannot be connected to generate a
curve. Indeed, since the $p_c^S$'s depend on both
$d$ and $q$, only the crosses corresponding to lattices with the same
topology, defined by the relation linking $d$ and $q$ can be connected. 

In practice, it means that the crosses corresponding to all the hypercubes
(sc) from
$d=3$ up to $d=13$ do belong to a single curve,
since the same relation $d=2q$ holds true for all these lattices. The
crosses corresponding to fcc lattices should also belong to another
curve, but the fcc percolation threshold for site is known for $d$=3, 4
and 5 only, and three data points are not sufficient to materialize a
curve. All the $p_c^S$'s up
to d=6 line up on the GM2 law (within the uncertainty limit above
mentioned) as stated in ref. \cite{gm}.

The new data for the sc lattices
at higher dimensions, however, give evidence of a deviation of the
$p_c^S$'s from the linear GM law which increases with $d$, illustrated in
Fig. (1). An equivalent formulation is to note a negative curvature of the
'numerical' $p_c^S(d)$ curve for sc hypercube lattices at $d>6$ to match
the  GM3 law. To quantify this effect, we have plotted in Fig. (2) the
relative difference $\Delta p_c^S/p_c^S=[p_c^S(nu)-p_c^S(GM2)]/p_c^S(nu)$
as a function of $d$ for the hypercubes (since data are available only for
these lattices at high dimensions).
$p_c^S(nu)$ is the numerical percolation
threshold
\cite{grass} and
$p_c^S(GM2)$ the prediction of the GM2 law.

We have reported in
\cite{gm} that
$|\Delta p_c^S|$ can reach 0.008 for some lattices in $d\le 6$.
This measures the accuracy of the GM2 law when applied to
{\sl any} Bravais lattice sc, bc, fcc in $3\le d\le 6$. However, regarding
the simple cubic and hypercubes only, the accuracy is
much better. In particular, in our prior work
(see Table 1 in ref. \cite{gm}), $|\Delta p_c^S|$ for sc lattices was
within
$5$x$10^{-4}$ at all dimensions $d<7$, except at $d=6$ where an
outstanding deviation $|\Delta p_c^S|=0.002$ was pointed out between the
numerical estimate 0.107 available at that time, and 0.109 predicted by
the GM2 law. The new numerical calculations
\cite{grass} have corrected the estimate of
$p_c^S$ at
$d=6$, raising $p_c^S$ from 0.107 to 0.109017, now in agreement with
the prediction of the GM2 law. With the new estimates of
ref.\cite{grass} which are $\simeq 30$ times more precise than the
previous ones, we find
$|\Delta p_c^S|/p_c^S\le 0.4\%$ in the whole range
$3\le d\le 6$ \cite{gm}. That is why we consider as significant a
departure from the GM2 law with $|\Delta p_c^S|/p_c^S\ge 0.4\%$ in Fig.
(2).

This plot then provides evidence for a
crossover at
$d_c=6$: at $d\le 6$, the GM2 law applies; at $d>6$, this is no longer
the case. The systematic quasi linear increase of $\Delta p_c^S/p_c^S$
as a function of $d$ which extrapolates to zero at $d=6$ corroborates
this value for the crossover dimension.
Note that $d_c$ is also known
to be the upper marginal dimension where critical exponents for the
percolation transition reach mean-field values. However, in phase
transition theory, only universal quantities
are supposed to undergo a crossover at $d_c$. This site percolation
transition gives an outstanding example where a non universal quantity
like $p_c^S$ also undergoes a crossover at
$d_c$.

Instead of choosing the GM2 law as the reference, we can also choose the
GM3 law, and investigate how the the percolation thresholds approach
this asymptotic law. The GM3
law is best illustrated in the log-log plot of
$2dp_c^S$ versus
$x=2d^2/[(d-1)(q-1)]$ in Fig. (3) since it
reduces to a straight line. For comparison, the data have been
also reported, and the crosses now correspond to the values of $p_c^S$ as
predicted by the GM2 law. As $x$ is not the pertinent variable
according to the GM2 law, once again, only the crosses
corresponding to all the hypercubes (sc) from
$d=3$ up to $d=13$ do belong to a single curve. As we can see
on fig. 2, this curve has a negative curvature and crosses the straight
line corresponding to the GM3 law. $p_c^S(nu)$ as a function of
$d$ then shifts from the GM2 law at $d_c=6$, to approach the GM3(law)
assumed to be its asymptote in the GM model.

To illustrate this behavior,
we have reported in Fig. (4)
$\Delta p_c^S/p_c^S=[p_c^S(nu)-p_c^S(GM3)]/p_c^S$ as a function of $d$
for hypercubes in high dimensions. From this Figure, it can be seen
that this asymptotic limit is indeed reached at $d=13$. It is then
important to note that the crossover at $d_c=6$ does not mean an abrupt
shift from the GM2 law to the GM3 law. Instead, it is a crossover to
another law which is missing here. More important, it indicates this law
$p_c^S(nu)$ as a function of $d$ is not embedded in the GM formula, and
accepts the GM3 law only as an asymptote in the large $d$ limit,
eventually reached (within the error bars) at
$d\ge 13$. Actually, the systematic and increasing deviation of $p_c^S$
from the GM3 law as $d$ decreases from $d=13$ can be viewed as
a pretransitional effect upon approaching the upper critical dimension
from below, beyond the scope of the GM3 law.

\subsection{Bond percolation}
Let us now investigate the situation for bonds. The GM2 law in this case
is illustrated in a log-log plot of $p_c^B$ as a function of
$(d-1)(q-1)/d$ (Fig. (5)). No deviation from the GM2 law can be evidenced
for any lattice up to the highest dimension $d=13$ investigated. To be
more specific, we have reported in Fig. (6) the differences $\Delta
p_c^B/p_c^B$ with $\Delta p_c^B = p_c^B(GM2)-p_c^B(nu)$ and $\Delta
p_c^B = p_c^B(nu)-p_c^B(GM3)$. Note the sign inversion in the
definition of $\Delta p_c^B$ to have this quantity positive, which means
that the numerical estimates are in between the estimates of the GM2 and
the GM3 laws.  Clearly, the exact (numerical)  results match the GM2 law
at
$d\le 6$, as expected. The relative deviation of
the GM2 law with respect to the numerical result at $d\le 6$ is in
the range
$\Delta p_c^B/p_c^B\le 0.9\%$, so that we can take as significant a
departure from the GM2 law $\Delta p_c^B/p_c^B\ge 0.9\%$. 

However, the
relative deviation between numerical estimates and the GM2 law is smaller
in all dimensions $7\le d\le 13$. Actually, the relative deviation of the
numerical estimates does not exceed $0.4\%$ in the whole range
$9\le d\le 13$, not only with respect to the GM2 law, but also with
respect to the GM3 law, so that no significant difference between the two
laws can be detected. Therefore, at contrast with the situation met in the
site percolation problem, the GM3 law is not accurate enough to detect a
 crossover at $d=6$. 
 
 To
explain why it is not possible to distinguish between the GM2 and
the GM3 laws in the range $7\le d\le 13$, we note that the leading term
in a
$q-1$ expansion of the GM2 law at large q is in $(q-1)^{b-2a}$. For bonds,
$b-2a = -0.94$, very close to the exponent -1 of the leading term in the
GM3 law. For sites, however, $b-2a=-1.23$ which is markedly different from
-1, so that a clear distinction between the GM2 law and the GM3 law can
be made even at the scale of a short dimension interval for sites, while
this is impossible for bonds. 

To be more specific, we note that the deviation of both the
GM2 and GM3 laws with respect to the numerical results, i.e.
$|p_c^B(GM2) - p_c^B(nu)|$ and $|p_c^B(GM3) - p_c^B(nu)|$, is only on the fourth digit 
in dimension $d>7$, which amounts to a relative error within $0.4\%$, while, as, we have 
stated above, a deviation with respect to the GM laws can be regarded as significant 
only if it exceeds $0.9\%$ for bonds. Therefore, the evidence of a crossover of dimensionality at $d_c=6$ for bonds requires an improvement of the GM law and its substitution by a new formula which would improve the accuracy typically a factor 3

\section{Discussion}
Van der Marck \cite{marc} has
shown that, if there is to be an exact universal formula for percolation
thresholds, it must be based on more information than $d$ and
$q$ only. In particular, the body centered cubic lattice and the stacked
triangular lattice both realize $d=3$, $q=8$, but they have different
$p_c$'s. This simple consideration is sufficient to show that Eqs. (1, 3)
cannot be exact. Nevertheless, the deviation with respect
to the exact or the almost exact (deduced from the best numerical
estimates) thresholds of sc, bcc and fcc lattices is so small
that the GM laws have the operational power to address the
problem of the crossover dimensionality.

On this basis, the present analysis of the percolation thresholds for sc
(hypercubes) hints at the  possible existence of a crossover at
$d_c=6$ for the site percolation threshold, although it is a 
non-universal parameter.
 At higher dimensions,
$p_c^S$ departs from the GM2 law to approach the GM3 asymptotic law which is
reached at
$d=13$. For bonds no similar behavior at $d_c=6$ is
detected.

Because of the relation $d=2q$ for hypercubes, the study of hypercubes
alone does not allow us to distinguish between the variables $d$ and $q$
to know which one is pertinent for the possible crossover.
However, some indication can be extracted from the data in Fig. (1). Let us
consider in particular the fcc lattice, which has a coordinance
$q$ much larger than the hypercube at the same dimension. For instance,
$q=24$ and
$40$ for fcc in dimensions
$d=4$ and
$d=5$, respectively. Those are the coordination numbers of hypercubes
in dimensions 12 and 20, respectively. The fcc
percolation thresholds at $d=4, 5$ have been reported in Fig. (1), along
with the values predicted by the GM2 and GM3 laws. This Figure shows
unambiguously that these two fcc lattices do satisfy the GM2 law and not
the GM3 law, despite the fact that their coordinance number are those
of the hypercubes in dimensions $d=12$ and $d=14$. 

It corroborates
that the pertinent variable responsible for the possible crossover 
behavior in the
percolation thresholds is indeed the dimension $d$, while
the coordination $q$ does not play any significant role. It also justifies
the term crossover of dimensionality used in this work. For universal
variables, such as the critical exponents, the renormalization group
technique provides us with an elaborate theory to understand the
crossover of dimensionality, at the associated identification of
universality classes. However such a behavior is not expected for 
 non-universal quantities like percolation thresholds. The explanation 
for this unexpected result is
thus a new challenge in the field of phase transition theory.

We note that $d_c$ is the upper critical limit for percolation transition
phenomena in general, not for the hypercubes only. The question then
arises whether the possible crossover of dimensionality we have observed for
hypercubes also applies to other lattices with a different topology. So
far, accurate numerical estimates of percolation
thresholds beyond $d=6$ could be achieved on the hypercubes only. The
lack of data prevents us for the moment from addressing
this question. Nevertheless, we believe that the progress in the
methods to compute percolation thresholds will make possible the
simulation of percolation of other systems more complex with a larger
coordinate number $q$ in the near future.

\begin{figure}
\begin{center}
\centerline{\epsfxsize=15cm\epsfbox{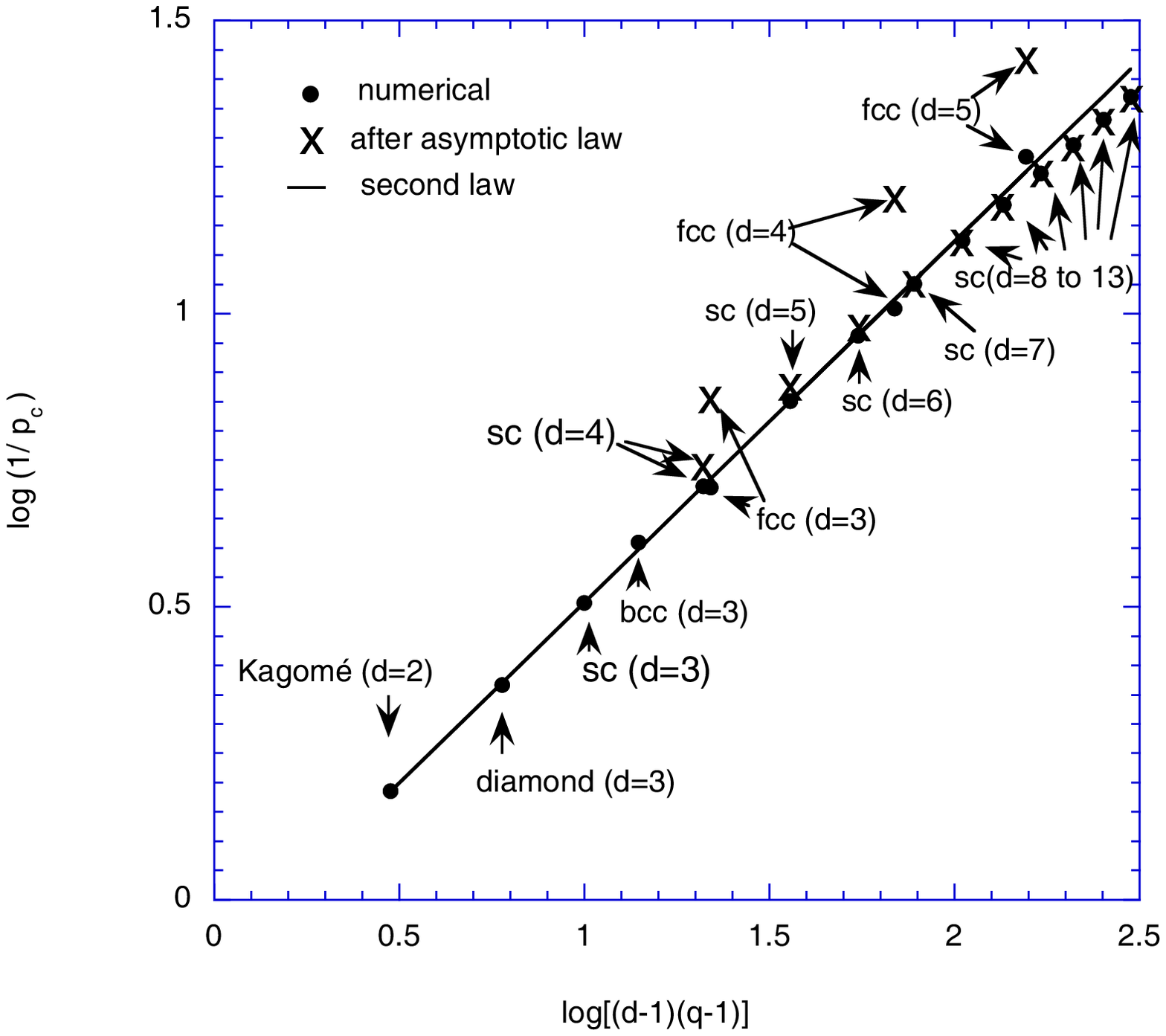}}
\caption{Site percolation threshold as a function of the variable
(d-1)(q-1) pertinent to the GM2 law, in decimal
logarithm. The numerical estimates ($\bullet$) are from
refs.\cite{grass,stau,gau}. The crosses are predictions of the GM3
asymptotic law, the solid line materializes the GM2 law.}
\end{center}
\end{figure}

\begin{figure}
\begin{center}
\centerline{\epsfxsize=15cm\epsfbox{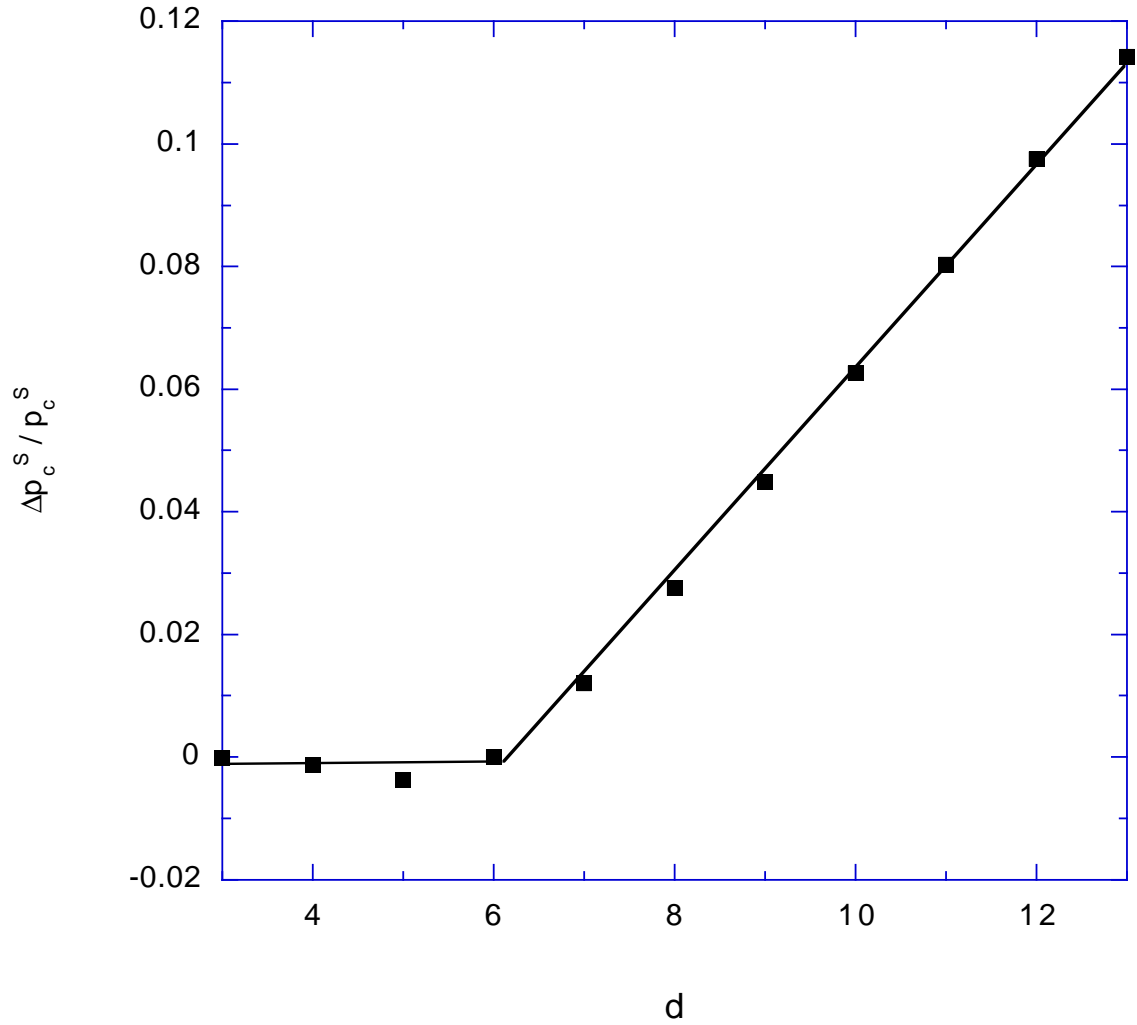}}
\caption{Relative difference $[p_c^S(nu)-p_c^S(GM2)]/p_c^S(nu)$ between
the numerical data $p_c^S(nu)$ and the the GM2 law $p_c^S(GM2)$ (full
squares) as a function of the dimension $d$ of the hypercubes. Lines are
guides for the eyes.}
\end{center}
\end{figure}

\begin{figure}
\begin{center}
\centerline{\epsfxsize=15cm\epsfbox{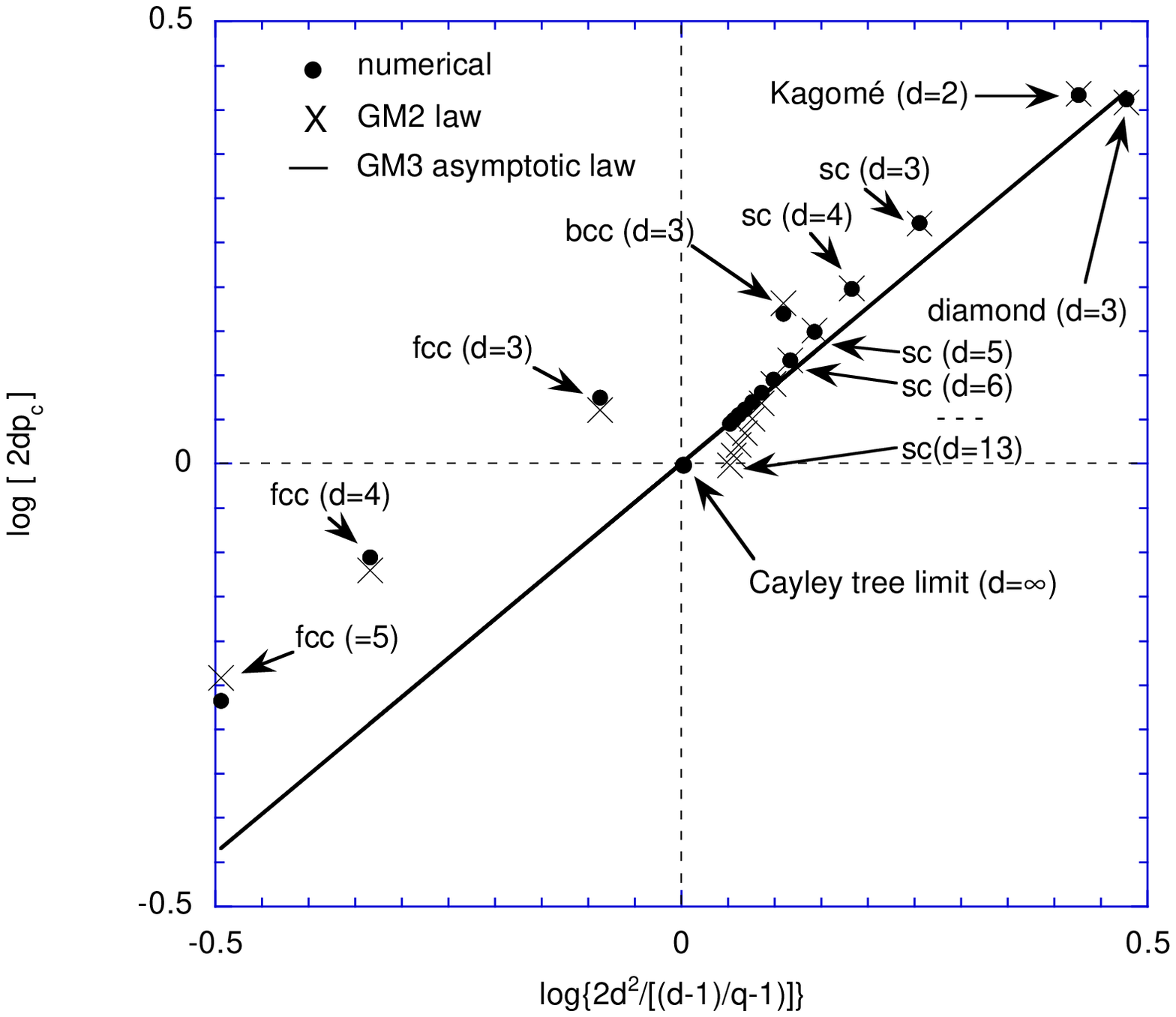}}
\caption{Site percolation threshold as a function of the variable
\mbox{$2d^2/[(d-1)(q-1)]$} pertinent to the GM3 law, in decimal
logarithm. The numerical estimates ($\bullet$) are from
refs.\cite{grass,stau,gau}. The crosses are predictions of the GM2
asymptotic law, the solid line materializes the GM3 law.
}
\end{center}
\end{figure}

\begin{figure}[t]
\begin{center}
\centerline{\epsfxsize=15cm\epsfbox{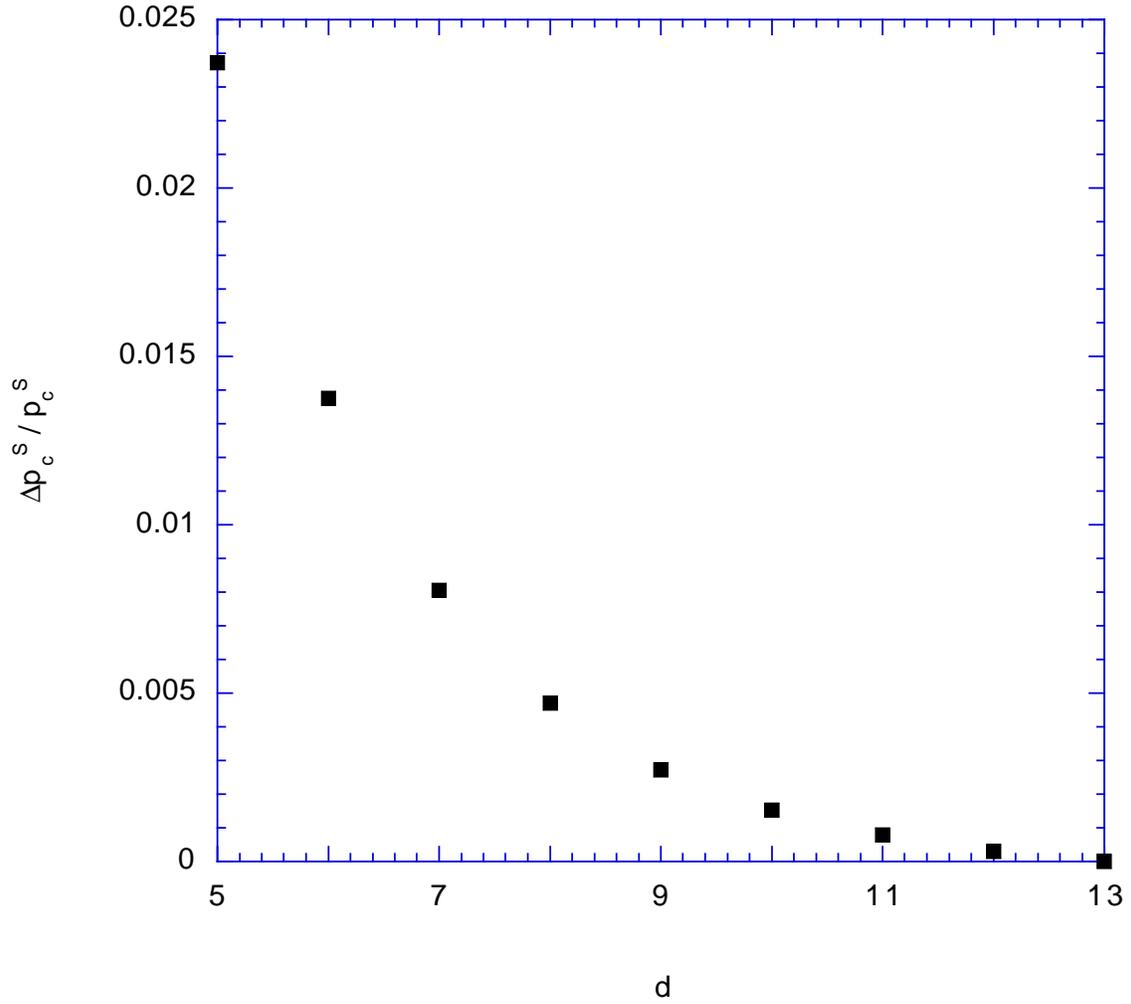}}
\caption{Relative
difference $[p_c^S(nu)-p_c^S(GM3)]/p_c^S(nu)$ between
the numerical data $p_c^S(nu)$and the the GM3 law $p_c^S(GM3)$
as a function of the dimension $d$ of the hypercubes.
}
\end{center}
\end{figure}

\begin{figure}
\begin{center}
\centerline{\epsfxsize=15cm\epsfbox{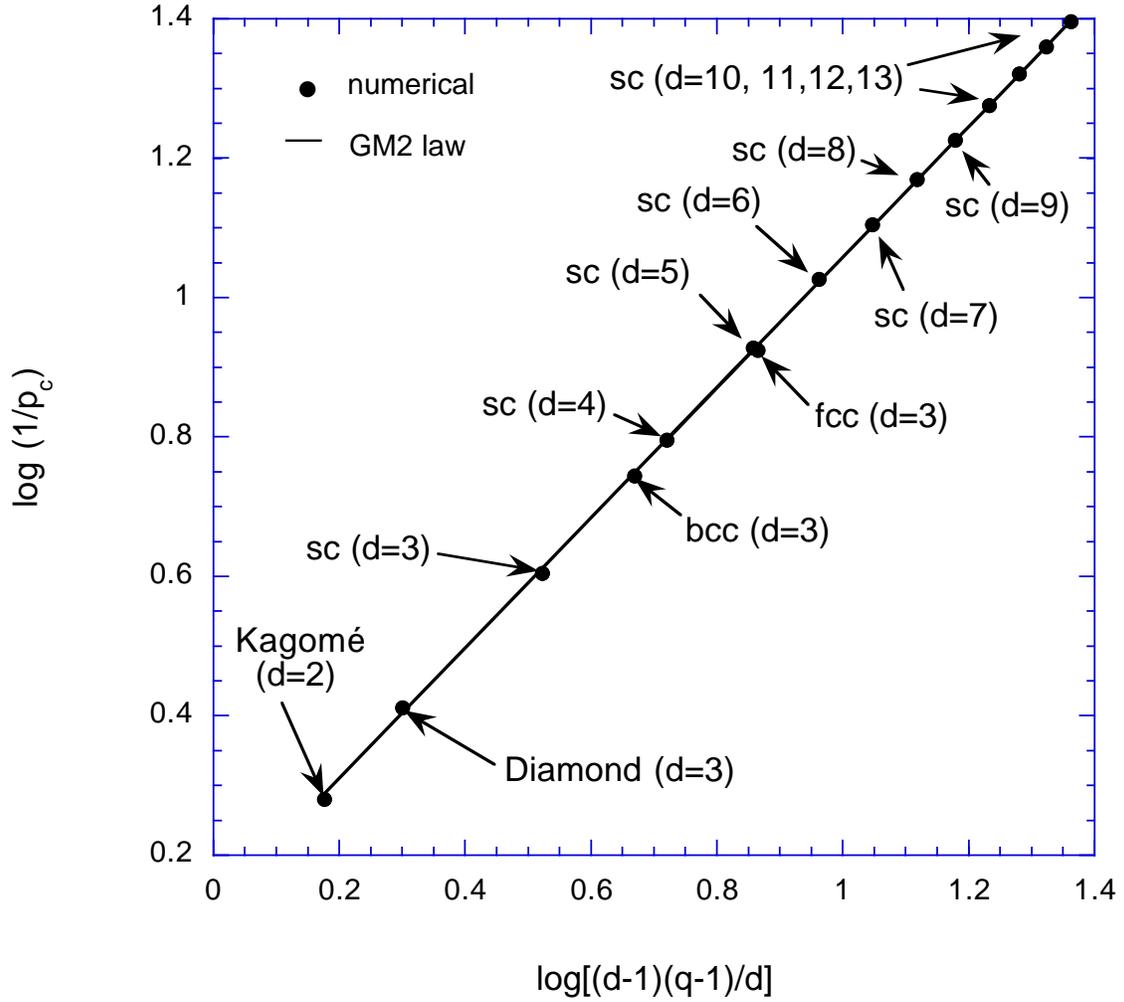}}
\caption{Bond percolation threshold as a function of the variable
\mbox{$(d-1)(q-1)/d$} pertinent to the GM2 law, in decimal
logarithm. The numerical estimates ($\bullet$) are from
refs.\cite{grass,stau,gau}. The solid line materializes the GM2 law.
}
\end{center}
\end{figure}

\begin{figure}
\begin{center}
\centerline{\epsfxsize=15cm\epsfbox{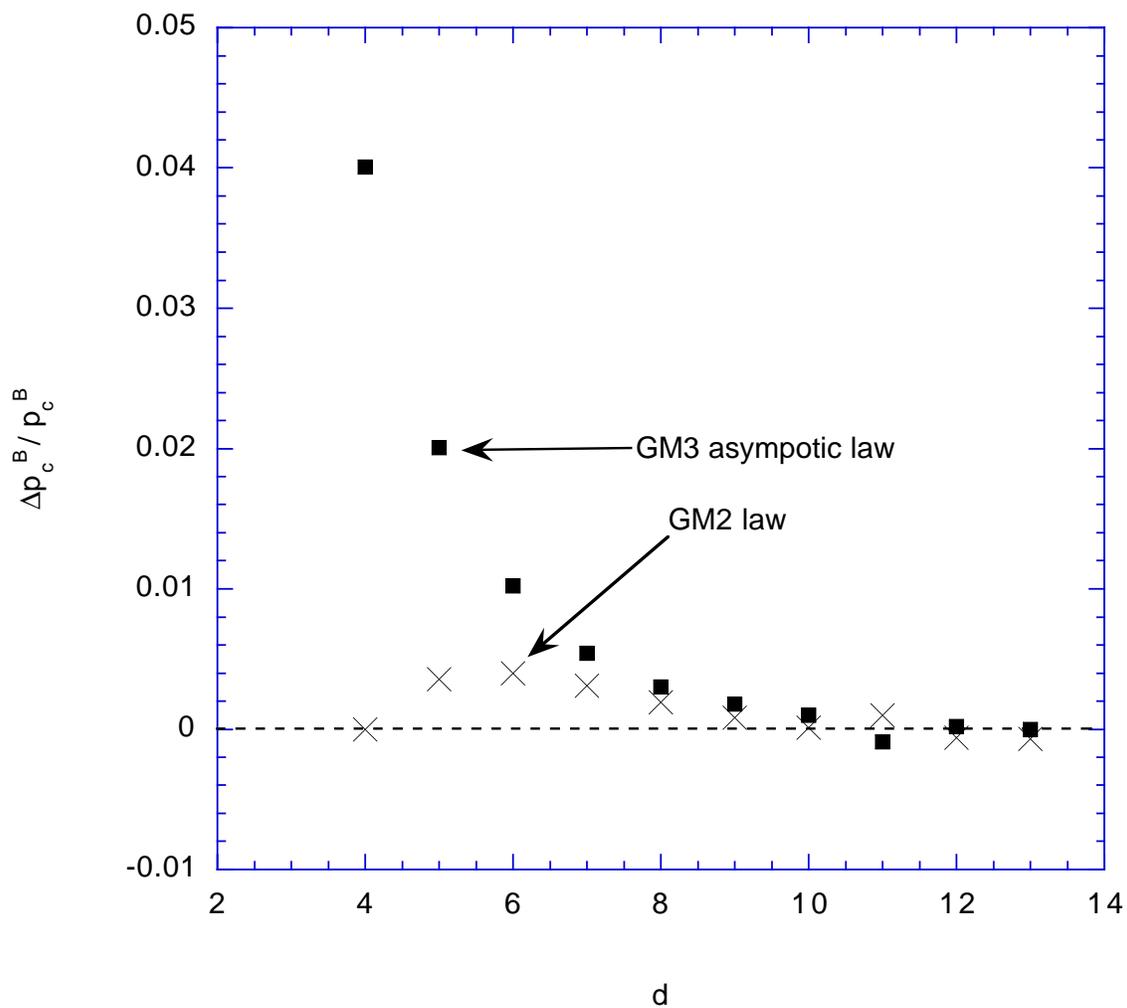}}
\caption{Relative difference $\Delta p_c^B/p_c^B(nu)$ with $\Delta p_c^B =
p_c^B(GM2)-p_c^B(nu)$ (crosses) and $\Delta p_c^B = p_c^B(nu)-p_c^B(GM3)$ 
( full squares) as a
function of the dimension $d$ of the hypercubes.
}
\end{center}
\end{figure}

\end{document}